\documentclass[11pt, oneside]{elsarticle}
\pdfoutput=1
\usepackage{lipsum}
\makeatletter
\def\ps@pprintTitle{%
 \let\@oddhead\@empty
 \let\@evenhead\@empty
 \def\@oddfoot{}%
 \let\@evenfoot\@oddfoot}
\makeatother
\RequirePackage[english=usenglishmax]{hyphsubst}
\usepackage{enumerate}
\usepackage{graphics}
\usepackage{url}
\usepackage{hyperref}	
\usepackage{breakurl}

\usepackage{float}
\usepackage{geometry}                		
\usepackage{amsmath}
\usepackage{ amssymb }
\usepackage{mathtools}
\usepackage{color}
\usepackage[table]{xcolor}
\usepackage{color,soul}
\usepackage{siunitx}
\usepackage{longtable}

\usepackage{booktabs}
\usepackage[flushleft]{threeparttable}

\usepackage{tikz,caption}
\newsavebox{\mytable}
\usetikzlibrary{fit,shapes.misc}

\usepackage{tabularx}

\usepackage{lmodern,multirow} 
\usepackage{rotating}

\newcolumntype{K}[1]{>{\centering\arraybackslash}p{#1}}

\usepackage{graphicx}				
\usepackage{hhline}

\begin{document}

\title{{Real implications of Quantitative Easing in the euro area: a complex-network perspective}}
\author[label1]{Chiara Perillo\corref{cor1}}
\ead{chiara.perillo@bf.uzh.ch}
\author[label1]{Stefano Battiston}
\ead{stefano.battiston@bf.uzh.ch}

\cortext[cor1]{Corresponding author}

\address[label1]{University of Zurich, Department of Banking and Finance, Zurich, Switzerland}

\begin{abstract}

The long-lasting socio-economic impact of the global financial crisis has questioned the adequacy of traditional tools in explaining periods of financial distress, as well as the adequacy of the existing policy response.  In particular, the effect of complex interconnections among financial institutions on financial stability has been widely recognized. A recent debate focused on the effects of unconventional policies aimed at achieving both price and financial stability. In particular, Quantitative Easing (QE, i.e., the large-scale asset purchase programme conducted by a central bank upon the creation of new money) has been recently implemented by the European Central Bank (ECB). In this context, two questions deserve more attention in the literature. First, to what extent, by injecting liquidity, the QE may alter the bank-firm lending level and stimulate the real economy. Second, to what extent the QE may also alter the pattern of intra-financial exposures among financial actors (including banks, investment funds, insurance corporations, and pension funds) and what are the implications in terms of financial stability. Here, we address these two questions by developing a methodology to map the macro-network of financial exposures among institutional sectors across financial instruments (e.g., equity, bonds, and loans) and we illustrate our approach on recently available data (i.e., data on loans and private and public securities purchased within the QE). We then test the effect of the implementation of ECB's QE on the time evolution of the financial linkages in the macro-network of the euro area, as well as the effect on macroeconomic variables, such as output and prices.

\end{abstract}


\maketitle

\section{Introduction}
\label{sec:1}

In the wake of the last wave of globalization, associated with the process of liberalization and financial integration, the last decades have witnessed the increase of financial interactions among the different actors of the global economy through different financial instruments. A large part of these interactions is intra-financial \cite{d2017compressing}, i.e., a great portion of banks' financial contracts has as a counterparty another bank or, more in general, another financial institution (e.g., investment funds, insurance corporations, and pension funds). As a consequence, before and during the global financial crisis, a large fraction of bank assets was intra-financial and this situation has continued after the crisis \cite{allahrakha2015systemic}. The 2007-2008 crisis has shown that these intra-financial linkages represent a mechanism for the propagation of financial distress and they may lead to the amplification of small shocks \cite{bardoscia2015debtrank, battiston2016rethinking}. Consequently, intra-financial interconnectedness is nowadays recognized as one of the key elements of potential financial instability or systemic risk (i.e., the risk of default of a large portion of the financial system) \cite{battiston2016financial,roukny2016interconnectedness}.

This fact, along with the long-lasting socio-economic impact of the financial crisis, has constituted the main impetus for a conversation about the adequacy of the existing tools in explaining and predicting periods of financial distress, as well as the adequacy of the existing policy response. In particular, a recent debate focused on the adoption, implementation, and effectiveness of unconventional policy tools, aimed at achieving both price and financial stability \cite{balasubramanyan2013bank, claessens2013macro, aiyar2014does, popoyan2017taming}. Among these unconventional policy tools, the expanded Asset Purchase Programme (APP) has been recently implemented by the European Central Bank (ECB). This large-scale purchase programme encompasses the set of measures that are commonly known as Quantitative Easing (QE). Indeed, the APP includes all purchase programmes under which private and public sector securities are purchased by the Eurosystem in an attempt to sustain inflation \cite{APP}. In particular, the APP, announced on January 2015, encompasses the existing i) third covered bond purchase programme (CBPP3) (started on 20 October 2014), and ii) asset-backed securities purchase programme (ABSPP) (started on 21 November 2014), combining them with the new iii) public sector purchase programme (PSPP) (started on 9 March 2015), and iv) corporate sector purchase programme (CSPP) (started on 8 June 2016) \cite{gambetti2017macroeconomic}. Monthly net purchases of public and private sector securities amount to 60 billion euros on average and they are intended to be carried out until the end of 2017, conditionally to a sustained adjustment in below, but close to, 2\% over the medium term \cite{APP}. The large majority of these purchases is made under the PSPP, which consists in the massive net acquisition of public sector securities by the Eurosystem from the banking system on the secondary market against central bank money \cite{ECB2015}.

The large-scale purchases imply the increase in the prices of the purchased assets and the provision of liquidity to the banking system. As a consequence, a wide range of interest rates fall and loans become cheaper \cite{QE}. Therefore, on the loans' demand side, firms and households may be able to borrow more and spend less to repay their debts, while, on the supply side, the availability of new liquidity may stimulate bank lending activity. Consequently, the QE may generate the increase in bank loans to the real economy. However, the QE does not necessarily imply that the money received by banks is going to be lent to the real economy. In fact, on the one hand, businesses and households could be not willing or able to borrow more from banks. On the other hand, banks could either decide to i) keep the money in the form of deposits at the Eurosystem (with low or even negative remuneration rate, but risk-free), or ii) invest it in the stock market (whose returns may be more appealing compared to the decreased returns of the securities purchased under the APP), as well as iii) increase the intra-financial interactions via different financial instruments (e.g., by increasing intra-financial loans). This may lead to the alteration of the pattern of intra-financial exposures among financial actors (including banks, investment funds, insurance corporations, and pension funds) with potential implications in terms of financial stability.

Despite there having been some efforts in the recent macro-econometric literature to investigate the effects of unconventional monetary measures, the exploration of the real implications of the recently implemented QE from a complex-network perspective would be of great importance. In particular, a financial network analysis would allow to account for the complex intra-financial interconnections and their potential effect on financial stability, neglected by traditional tools. Network theory has been widely applied in the financial domain to provide useful insights and predictions on the interconnectivity of the financial system and on the propagation of shocks, especially with reference to the banking sector \cite{eisenberg2001systemic, elsinger2006risk, battiston2012debtrank, roukny2013default, battiston2016leveraging}. However, less attention has been devoted to the study of the interplay between monetary policy, macroeconomic variables, and network of financial exposures, able to take into account both the interactions between the real and financial sector and intra-financial interconnectedness. In particular, in the light of the above context, two fundamental questions deserve more attention in the literature:

\begin{enumerate}
\item {To what extent, by injecting liquidity, does the QE alter the bank-firm lending level and stimulate the real economy?}
\item {To what extent, does the QE also alter the pattern of intra-financial exposures among financial actors and what are the implications in terms of financial stability?}
\end{enumerate}

Here, we address these two questions and investigate the real implications of ECB's QE, by taking a complex-network perspective to unconventional monetary policy. In particular, first, we develop a methodology to map the macro-network of financial exposures among the different institutional sectors operating in the euro area (i.e., \textit{Monetary Financial Institutions}, \textit{Insurance Corporations and Pension Funds}, \textit{Financial Corporations excluding Monetary Financial Institutions and Insurance Corporations and Pension Funds}, \textit{Households and Non-Profit Institutions Serving Households}, \textit{Non-Financial Corporations}, and \textit{General Governments}). This macro-network can be regarded as a multiplex weighted network in which multiple types of links correspond to different financial instruments (e.g., equity holdings, corporate and sovereign bonds, and loans). Second, we illustrate our approach on recently available data, with a particular focus on i) loans granted by banks to the institutional sectors operating in the euro area and ii) private and public securities purchased by the Eurosystem from euro area banks within the APP. Third, we investigate the time evolution of the financial linkages in the macro-network of the euro area, as well as the time evolution of macroeconomic variables, such as output and prices, since the implementation of ECB's QE. In particular, in an attempt to assess the transmission of this unconventional monetary measure to the real economy through the bank-lending channel (which focuses on the quantity of loans supplied by banks), we estimate the time evolution of the exposure of the banking system to the real economy through loans. Further, we investigate the intra-financial interactions in the euro area, by focusing on the estimate of the time evolution of the banking system's exposure to the financial sector via loans, since the initiation of the APP. Finally, we assess the effects of QE on the time evolution of \textit{Gross Domestic Product (GDP)} and \textit{Harmonised Index of Consumer Prices (HICP)}. The combination of network variables (i.e., time series of financial exposures) and macroeconomic variables allows us to shed light on the implications of QE both in terms of stimulation of the real economy and intra-financial interconnectedness. This approach represents a tangible step ahead in the comprehensive analysis of the effects of unconventional monetary policy that until now involved macroeconomic implications or the impact on financial markets, neglecting network effects and their interplay with macroeconomic variables.

The remainder of this paper is organized as follows. In Section \ref{sec:methodology}, we introduce the methodology that we developed to map the macro-network of financial exposures among the different institutional sectors operating in the euro area and we illustrate the employed data and sources. In Section  \ref{sec:exploratory}, we illustrate the exploratory results both from a static and a dynamic perspective. Finally, Section \ref{sec:conclusions} concludes the paper.


\section{Methodology and data}
\label{sec:methodology}

\subsection{Institutional sectors}

We consider the euro area economy as composed by three main macro sectors: i) the financial sector, ii) the non-financial private sector (or the real sector), and iii) the non-financial public sector. The financial sector includes the following institutional sectors: i) \textit{Monetary Financial Institutions (MFI)}, ii) \textit{Insurance Corporations and Pension Funds (IC\&PF)}, iii) \textit{Financial Corporations excluding MFI and IC\&PF (FC excl. MFI and IC\&PF)}. More specifically, \textit{Monetary Financial Institutions} encompass i) the \textit{Eurosystem}, i.e., the European Central Bank and the National Central Banks \textit{(ECB\&NCB)} of the nineteen countries that have adopted the euro in stage three of Economic and Monetary Union (EMU)\footnote{The European Union's Member States that have adopted the euro in stage three of EMU are Austria, Belgium, Cyprus, Estonia, Finland, France, Germany, Greece, Ireland, Italy, Latvia, Lithuania, Luxembourg, Malta, Netherlands, Portugal, Slovakia, Slovenia, and Spain.}, and ii) \textit{Monetary Financial Institutions excluding the Eurosystem (MFI excl. ECB\&NCB)}, broadly equivalent to banks and money market funds. Further, \textit{Financial Corporations excluding MFI and IC\&PF} include i) investment banks other than money market funds, ii) other financial intermediaries different from \textit{IC\&PF}, iii) financial auxiliaries, and iv) captive financial institutions and money lenders. As regards the real sector, it encompasses the institutional sectors i) \textit{Households and Non-Profit Institutions Serving Households (HH\&NPISH)}, and ii) \textit{Non-Financial Corporations (NFC)}. Finally, the non-financial public sector could be identified with \textit{General Governments (GG)}, which comprise central, state (regional) and local government and social security funds \cite{InstitutionalSectors}. The macro sectors considered in our study and their breakdown by institutional sector with relative acronyms are summarized in Table \ref{tab:1}.


\begin{table}
\begin{center}
\caption{The euro area macro sectors and their breakdown by institutional sector}
\label{tab:1}
\renewcommand{\arraystretch}{1.1}
\begin{tabular}[h!]{|p{0.3\textwidth}|p{0.3\textwidth}|p{0.3\textwidth}|}
\hline
\textbf{Macro sectors} & \textbf{Institutional sectors} & \textbf{Acronym} \\
\hline \\[-2.8ex]\hline
Financial sector & Monetary Financial Institutions (Eurosystem $+$ Monetary Financial Institutions excluding the Eurosystem)   &  MFI (ECB\&NCB $+$ MFI excl. ECB\&NCB) \\
\hhline{~--}
 & Insurance Corporations and Pension Funds  &  IC\&PF \\ 
\hhline{~--}
  & Financial Corporations excluding MFI and IC\&PF  &  FC excl. MFI and IC\&PF \\ 
\hline
Non-financial private sector (or real sector)        & Households and Non-Profit Institutions Serving Households   &  HH\&NPISH\\ 
\hhline{~--}
  & Non-Financial Corporations       &  NFC \\ 
\hline
Non-financial public sector  & General Governments  &  GG\\ 
\hline
\end{tabular}
\end{center}
\end{table}

\subsection{Methodology}

We regard the euro area economy as a macro-network of financial exposures, where nodes are the aforementioned institutional sectors (summarized in Table \ref{tab:1}). Despite the fact that every node of the macro-network is represented at an aggregate level, it is worth mentioning that each node is the result of an aggregation on two dimensions: by country and by sector. More specifically, the euro area macro-network of financial exposures includes the nineteen macro-networks of financial exposures among the institutional sectors operating in the nineteen European Union's Member States belonging to the Eurozone and their mutual interactions. Further, every node of the macro-network should be regarded as a network of micro-based granular financial exposures among the different institutions operating within every institutional sector. The euro area macro-network of financial exposures can be regarded as a multiplex weighted network in which multiple types of links correspond to different financial instruments (e.g., loans, corporate and sovereign bonds, and equity). Links follow the direction of the money and weights correspond to the balance sheet outstanding amount at the end of the period (stocks) of the financial exposures existing among the different institutional sectors through the different financial instruments. 

\subsection{Variables}

The empirical analysis has been conducted with regard to the period spanning from the first quarter of 2003 to the second quarter of 2017 (2003Q1-2017Q2, both quarters included), considering six institutional sectors operating in the euro area (i.e., \textit{Monetary Financial Institutions (MFI)}, composed by \textit{Eurosystem (ECB\&NCB)} and \textit{Monetary Financial Institutions excluding the Eurosystem (MFI excl. ECB\&NCB)}, \textit{Insurance Corporations and Pension Funds (IC\&PF)}, \textit{Financial Corporations excluding MFI and IC\&PF (FC excl. MFI and IC\&PF)}, \textit{Households and Non-Profit Institutions Serving Households (HH\&NPISH)}, \textit{Non-Financial Corporations (NFC)}, and \textit{General Governments (GG)}). Our study focuses on two categories of variables: i) network variables and ii) macroeconomic variables. We define network variables as the time series of financial exposures through which institutional sectors are tied to each other. In particular, our work focuses on the time evolution of the financial link between the Eurosystem and the banking system via \textit{APP securities}, existing as a consequence of the APP purchases of public and private securities that imply an injection of money into the banking system. Furthermore, we focus on the time series of the financial dependencies via loans between the euro area banking sector and financial sector \textit{(Loans from MFI excl. ECB\&NCB to MFI, Loans from MFI excl. ECB\&NCB to IC\&PF, Loans from MFI excl. ECB\&NCB to FC excl. MFI and IC\&PF)}, real sector \textit{(Loans from MFI excl. ECB\&NCB to HH\&NPISH, Loans from MFI excl. ECB\&NCB to NFC)}, and public sector \textit{(Loans from MFI excl. ECB\&NCB to GG)}. As regards the macroeconomic variables, our study focuses on the time evolution of the euro area \textit{Gross Domestic Product (GDP)} and \textit{Harmonised Index of Consumer Prices (HICP)}.

Regarding the variable \textit{APP securities}, we consider the quarterly data derived as end-of-the quarter monthly data of the Eurosystem APP holdings and a breakdown in the different programmes included in the APP: i) third covered bond purchase programme (CBPP3), ii) asset-backed securities purchase programme (ABSPP), iii) public sector purchase programme (PSPP), and iv) corporate sector purchase programme (CSPP). The time series of the cumulative purchase breakdowns under the APP have been obtained from \cite{APP}\footnote{Please note that all data sources in this paper have been last accessed on 19 July 2017.}. The APP holdings refer to private and public sector securities purchased by the Eurosystem within the APP in order to address the risk of a too prolonged period of low inflation \cite{APP}. In particular, the purchased securities include: i) covered bonds (purchased under the CBPP3), ii) asset-backed securities (purchased under the ABSPP), iii) nominal and inflation-linked central government bonds (purchased under the PSPP), iv) bonds issued by recognized agencies, regional and local governments, international organizations, and multilateral development banks located in the euro area (purchased under the PSPP), and v) corporate sector bonds (purchased under the CSPP) \cite{APP}. These purchases increase the Eurosystem holdings of the aforementioned marketable debt instruments and particularly the purchases made within the CBPP3, ABSPP, and PSPP imply an injection of liquidity into the banking system \cite{APP}.

As regards the loans from the banking sector to the financial sector, the real sector and the public sector, the time series of the variables \textit{Loans from MFI excl. ECB\&NCB to MFI, Loans from MFI excl. ECB\&NCB to IC\&PF, Loans from MFI excl. ECB\&NCB to FC excl. MFI and IC\&PF, Loans from MFI excl. ECB\&NCB to HH\&NPISH\footnote{Loans from \textit{MFI excl. ECB\&NCB} to households include: i) credit for consumption, ii) lending for house purchase, and iii) other lending. The time series of the breakdowns of loans into the three aforementioned categories can be taken from \cite{ECBStatistical}.}, Loans from MFI excl. ECB\&NCB to NFC,} and \textit{Loans from MFI excl. ECB\&NCB to GG} have been obtained from ECB Statistical Data Warehouse \cite{ECBStatistical}. The aforementioned variables are derived as end-of-quarter monthly data of the balance sheet outstanding amounts of loans granted by \textit{MFI excl. ECB\&NCB} to the different institutional sectors operating in the euro area and the resulting series have been seasonally and working day adjusted. 

Finally, regarding the macroeconomic variables, \textit{Gross Domestic Product (GDP)} at market prices and \textit{Harmonised Consumer Price Index (HICP)} (2015=100) have been obtained from \cite{ECBStatistical}. According to the European System of National and Regional Accounts (ESA 2010), the \textit{GDP} at market prices can be defined as the final result of the production activity of resident producer units \cite{DefinitionGDPmarketprices}, while the \textit{HICP} is a ``Laspeyres-type price index", defined in terms of household final monetary consumption expenditure \cite{HICPsGuide}. The \textit{GDP} series is chain-linked volume and the \textit{HICP} variable is derived as end-of-quarter monthly data. Both \textit{GDP} and \textit{HICP} time series are working day and seasonally adjusted.

\section{Exploratory results}
\label{sec:exploratory}

\begin{figure}[t]
\par\medskip
\centering
\includegraphics[scale=0.57]{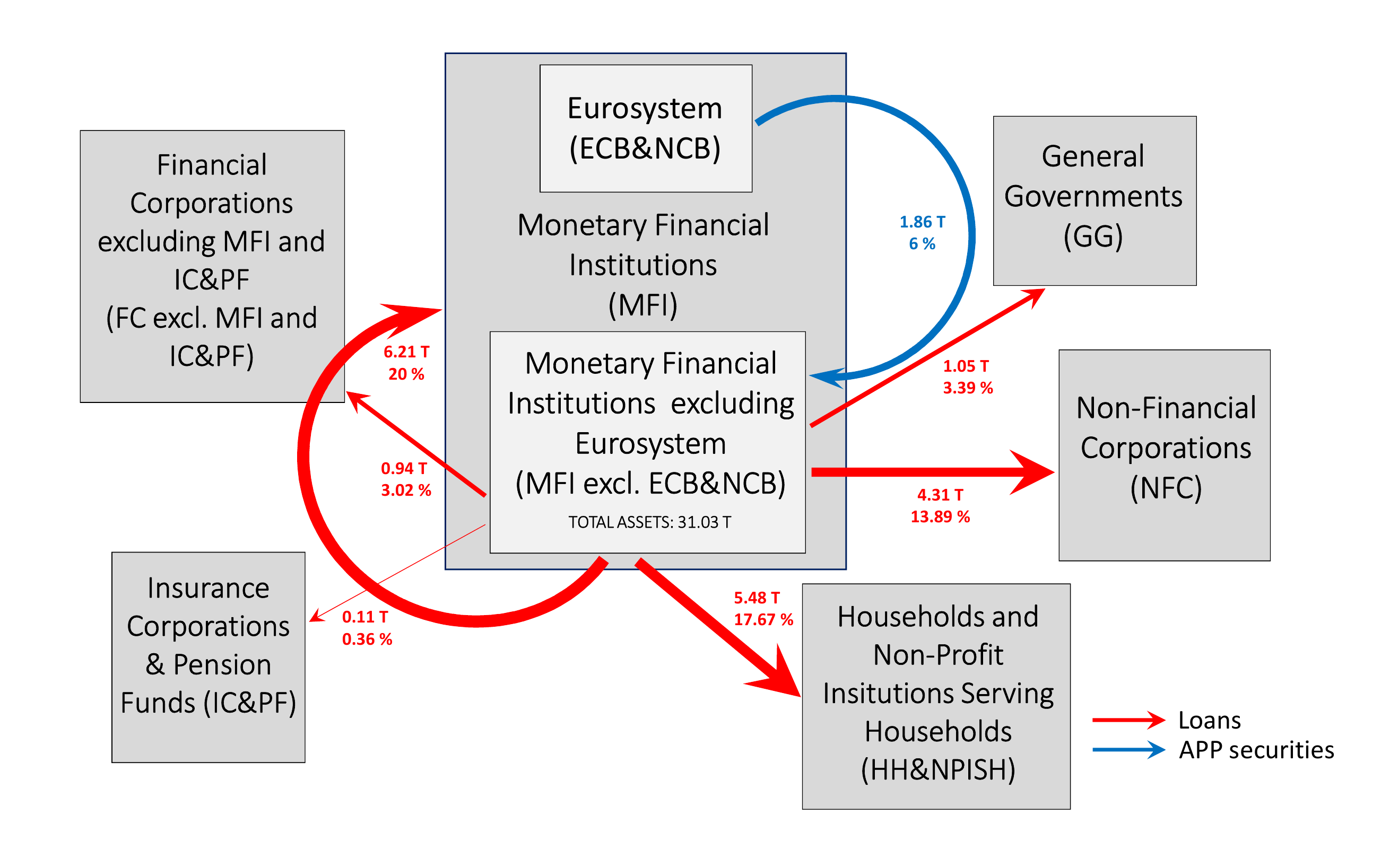}
\caption{The macro-network of financial exposures in the euro area - June 2017. Macro-network of financial exposures among the different institutional sectors operating in the euro area (i.e., \textit{Monetary Financial Institutions (MFI)}, composed by \textit{Eurosystem (ECB\&NCB)} and \textit{Monetary Financial Institutions excluding the Eurosystem (MFI excl. ECB\&NCB)}, \textit{Insurance Corporations and Pension Funds (IC\&PF)}, \textit{Financial Corporations excluding MFI and IC\&PF (FC excl. MFI and IC\&PF)}, \textit{Households and Non-Profit Institutions Serving Households (HH\&NPISH)}, \textit{Non-Financial Corporations (NFC)}, and \textit{General Governments (GG)}), via i) private and public securities purchased by the Eurosystem from the banking system within the APP (blue link) and ii) loans granted from the banking sector to the institutional sectors operating in the euro area (red links), with reference to the second quarter of 2017. Data source: \cite{ECBStatistical}.}
\label{Fig2}
\end{figure}

In an attempt to investigate i) to what extent the QE may alter the bank-firm lending level and stimulate the real economy and ii) to what extent it may also alter the intra-financial interactions, for example, via the increase of intra-financial loans, first, we map the euro area macro-network of financial exposures. Figure \ref{Fig2} illustrates the macro-network of financial exposures among the different institutional sectors operating in the euro area, with reference to the second quarter of 2017. In particular, our main focus is on two financial instruments: i) the private and public securities purchased by the Eurosystem from the banking system within the APP (represented with a blue arrow) and ii) the loans granted from the banking sector to the institutional sectors operating in the euro area (represented with red arrows). As it can be seen from Figure \ref{Fig2}, the arrows follow the direction of the money. The numbers correspond to the balance sheet outstanding amount at the end of the period (stocks) of i) APP holdings of the Eurosystem related to the asset purchases made within the CBPP3, ABSPP, and PSPP and ii) loans granted by the euro area \textit{MFI excl. ECB\&NCB}. Moreover, in order to give a clearer idea of the weight of every link in terms of asset size of the banking system, we also express loan and APP exposures as a percentage of total assets of the euro area banking system.

As shown in Figure \ref{Fig2}, the injection of money by the Eurosystem to the banking sector (blue arrow) corresponds to 6\% of the total assets of the euro area \textit{MFI excl. ECB\&NCB}. As regards loans (red links), it is worth mentioning that, at the end of June 2017, 20\% of the euro area banks' assets were bank-to-bank loans, while around 18\% were loans to households, followed by bank-firm loans, corresponding approximatively to 14\% of banks' assets. Therefore, from the macro-network analysis we can notice that, despite the fact that a large fraction of loans is intra-financial (i.e., 23.38\% of banks' total assets), a relevant fraction of loans is granted by the banking sector to the real sector (corresponding to 31.56\% of banks' total assets). However, the macro-network methodology provides us only with a static picture of the financial dependencies existing between the euro area banking sector and the other euro area institutional sectors. For this reason, we decided to investigate the time evolution of the financial linkages in the macro-network of the euro area, as well as the time evolution of the macroeconomic variables output \textit{(GDP)} and prices \textit{(HICP)}, since the implementation of ECB's QE, able to provide us with further insights.

\begin{figure}[H]
\par\medskip
\centering
\includegraphics[scale=0.75]{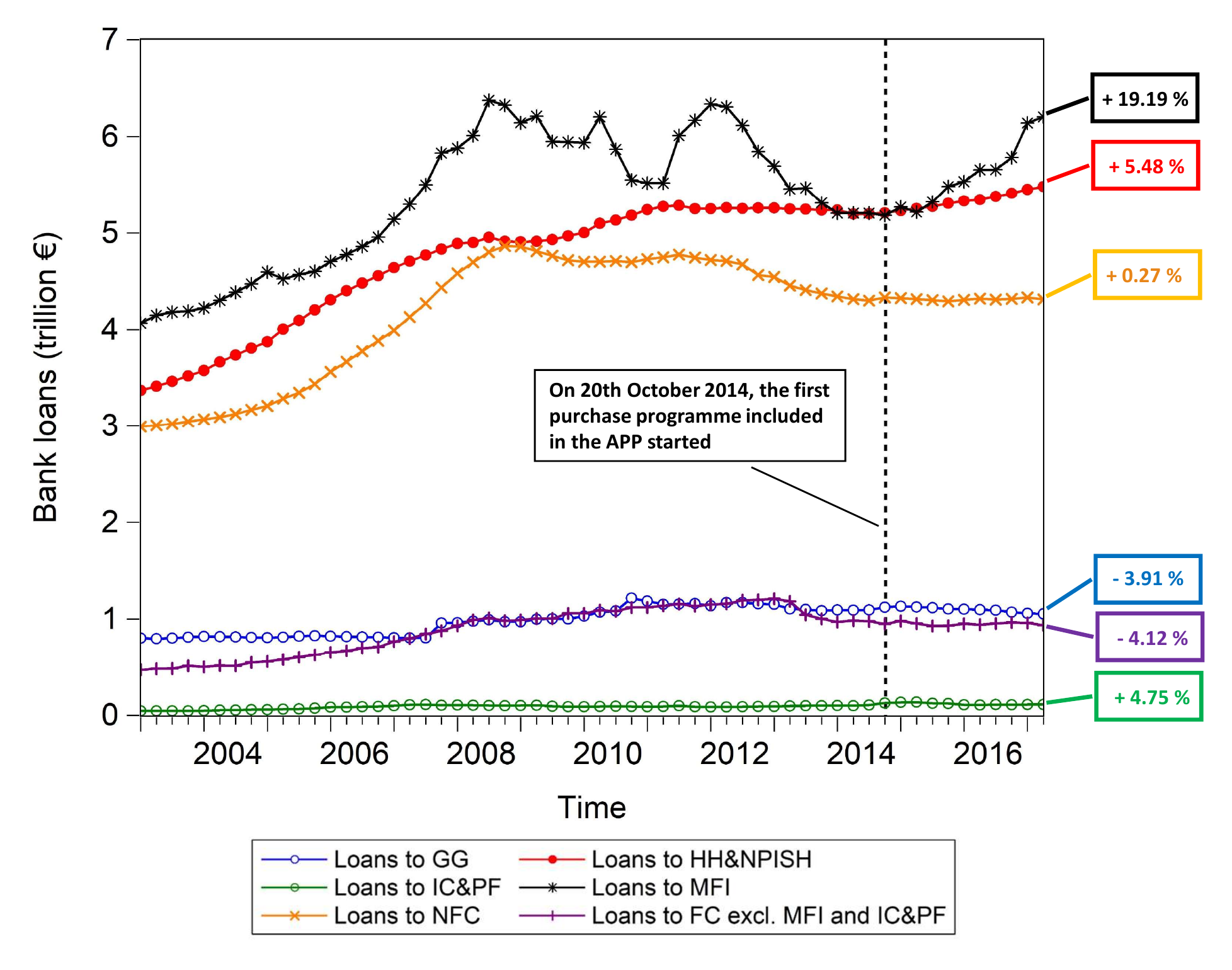}
\caption{Euro area bank loans - breakdown by institutional sector. Time series over the reference period 2003Q1-2017Q2 of the breakdown of loans granted by euro area banks to: i) \textit{Monetary Financial Institutions (MFI}, in black), ii) \textit{Households and Non-Profit Institutions Serving Households (HH\&NPISH}, in red), iii) \textit{Non-Financial Corporations (NFC}, in orange), iv) \textit{General Governments (GG}, in blue), v) \textit{Insurance Corporations and Pension Funds (IC\&PF}, in green), and vi) \textit{Financial Corporations excluding MFI and IC\&PF (FC excl. MFI and IC\&PF}, in purple). The vertical dashed line corresponds to the beginning of the first purchase programme (third covered bond purchase programme, CBPP3) included in the APP, started on 20 October 2014, while the percentages on the right-hand side of the graph correspond to our estimate of the growth rates of loans granted by banks to the different institutional sectors of the euro area, since the initiation of QE. Data source: \cite{ECBStatistical}.}
\label{Fig3}
\end{figure}

In particular, Figure \ref{Fig3} represents the time evolution of loans granted by banks to the different institutional sectors operating in the euro area, over the reference period 2003Q1-2017Q2. As it can be seen from Figure \ref{Fig3}, the time series of bank-to-bank loans is represented in black, while the time series of loans granted from the banking sector to households is represented in red, and the time series of bank-firm loans is represented in orange. Further, the time series of loans granted by the banking sector to \textit{General Governments}, \textit{Financial Corporations excl. MFI and IC\&PF}, and \textit{Insurance Corporations and Pension Funds} are represented in blue, purple, and green, respectively. The vertical dashed line corresponds to the beginning of the first purchase programme included in the APP, started on 20 October 2014. On the right-hand side of Figure \ref{Fig3} we reported our estimate of the growth rate of loans granted by banks to the different institutional sectors of the euro area, since the initiation of QE. Regarding the loans' growth rates, it is worth mentioning that, since QE has been initiated, bank-to-bank loans have increased by 19.19\%, while loans to households have grown by 5.48\%, and bank-firm loans only by 0.27\%. Consequently, the empirical evidence provided by data on loans allows us to state that, since QE, there has been an increase in bank lending activity at an aggregate level, but mostly addressed to the banking system itself.

\begin{figure} [H]
\par\medskip
\centering
\includegraphics[scale=0.70]{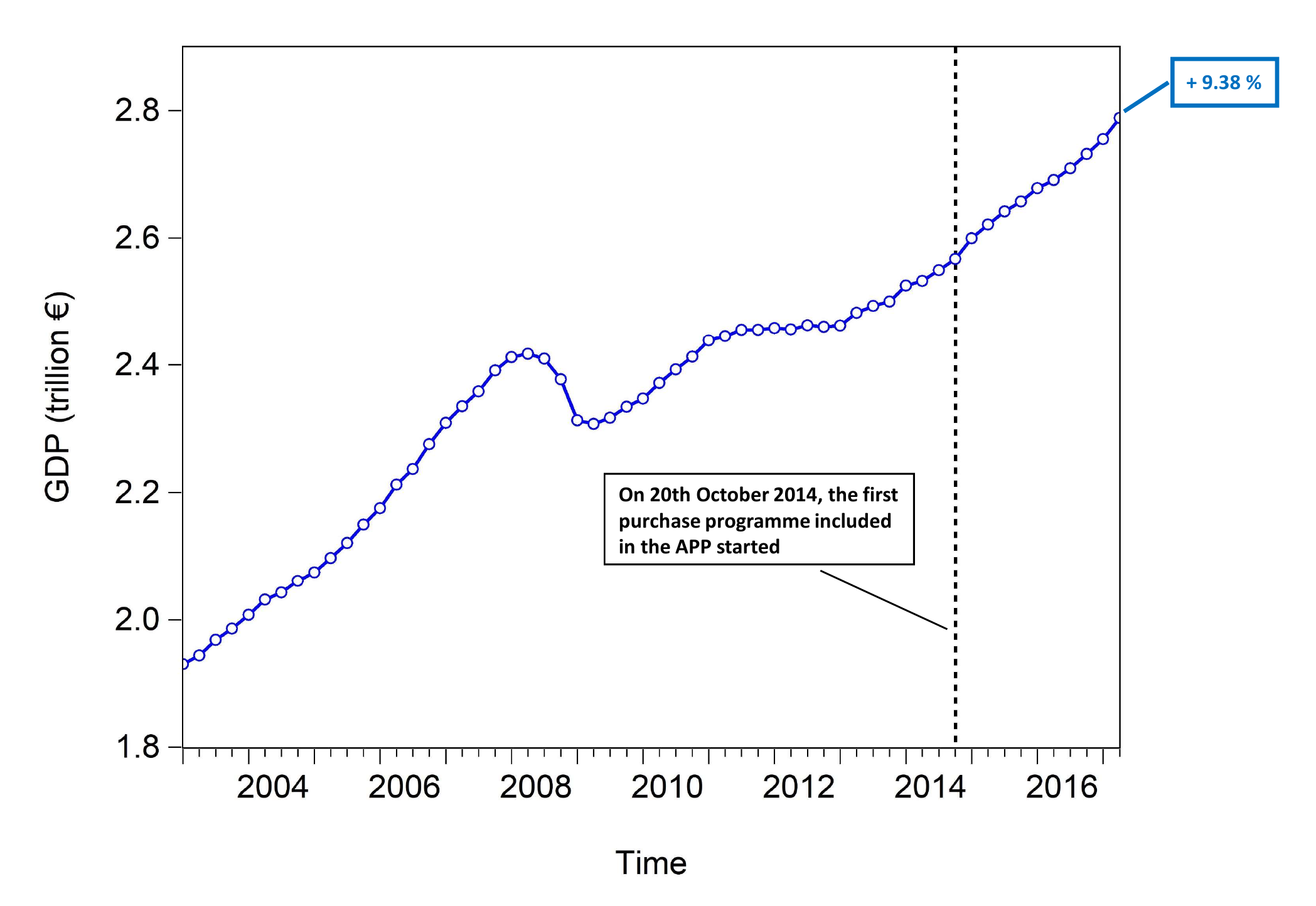}
\caption{Euro area GDP. Time series of the euro area \textit{Gross Domestic Product (GDP)} at market prices, over the reference period 2003Q1-2017Q2. The vertical dashed line corresponds to the beginning of the first purchase programme (third covered bond purchase programme, CBPP3) included in the APP, started on 20 October 2014, while the percentage on the right-hand side of the graph corresponds to our estimate of the growth rate of the euro area \textit{GDP}, since the initiation of QE. Data source: \cite{ECBStatistical}.}
\label{Fig4}
\end{figure}

\begin{figure} [H]
\par\medskip
\centering
\includegraphics[scale=0.70]{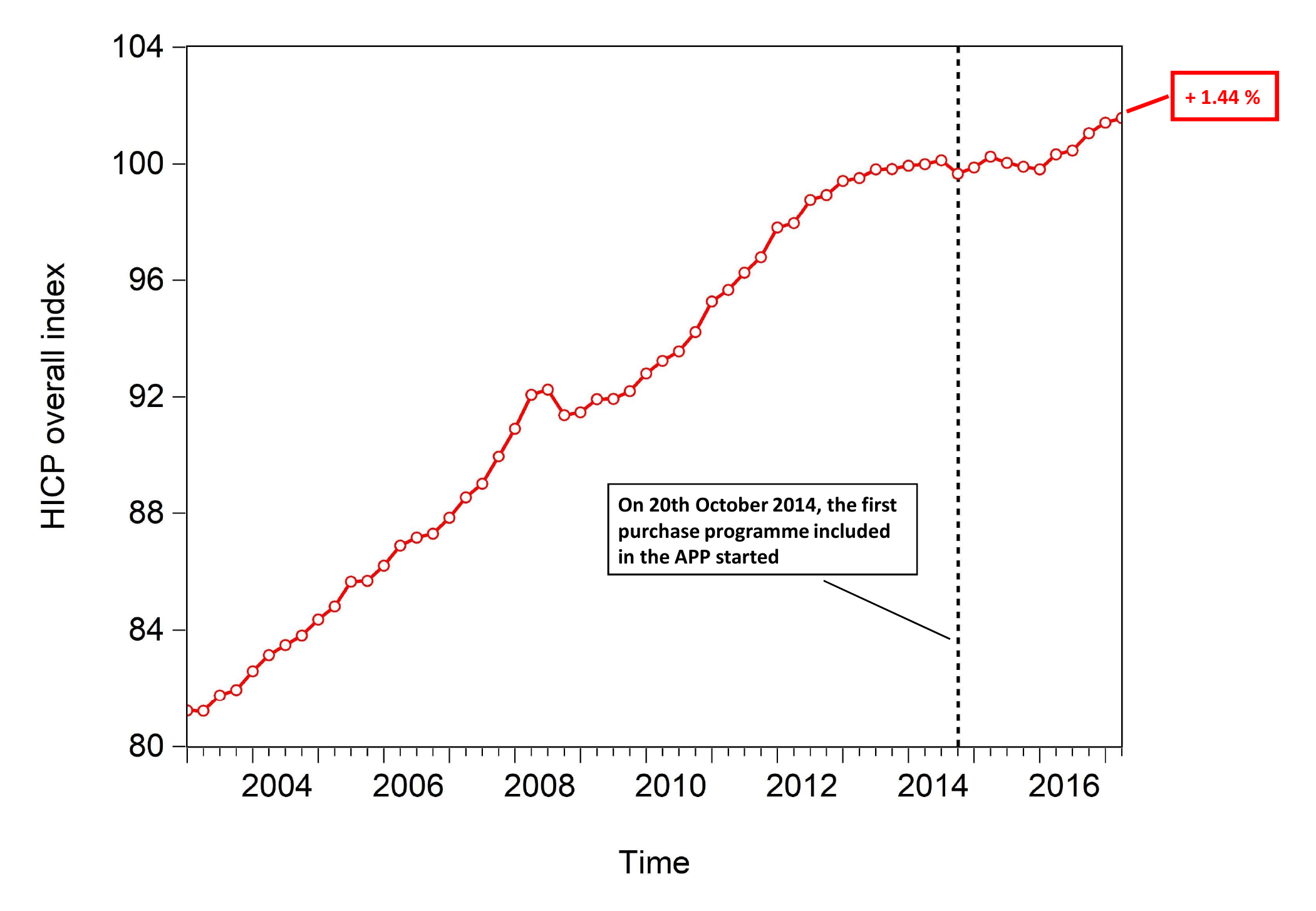}
\caption{Euro area HICP. Time series of the \textit{Harmonised Consumer Price Index (HICP)}, over the reference period 2003Q1-2017Q2. The vertical dashed line corresponds to the beginning of the first purchase programme (third covered bond purchase programme, CBPP3) included in the APP, started on 20 October 2014, while the percentage on the right-hand side of the graph corresponds to our estimate of the growth rate of the euro area \textit{HICP}, since the initiation of QE. Data source: \cite{ECBStatistical}.}
\label{Fig5}
\end{figure}

Further, Figures \ref{Fig4} and \ref{Fig5} represent the time evolution of the euro area macroeconomic variables \textit{Gross Domestic Product (GDP)} at market prices and \textit{Harmonised Consumer Price Index (HICP)}, respectively, over the reference period 2003Q1-2017Q2. As it can be seen from Figures \ref{Fig4} and \ref{Fig5}, the time series of \textit{GDP} is represented in blue, while the time series of \textit{HICP} is represented in red. The vertical dashed line corresponds to the beginning of the first purchase programme included in the APP, started on 20 October 2014. On the right-hand side of Figures \ref{Fig4} and \ref{Fig5} we reported our estimate of the growth rate of both \textit{GDP} and \textit{HICP}, since the initiation of QE. In this regard, it is worth noting that, since the APP programme has been implemented, both \textit{GDP} and inflation have increased. In particular, the \textit{GDP} has grown by 9.38\%, while the \textit{HICP} has increased by 1.44\%.

\section{Conclusions}
\label{sec:conclusions}

In this paper, we have conducted an empirical explorative analysis to study the real implications of ECB's QE in an attempt to investigate i) to what extent the QE may alter the bank-firm lending level and stimulate the real economy and ii) to what extent it may also alter the intra-financial interactions, for example, via the increase of intra-financial loans. Our empirical results constitute a contribution to the debate on the effectiveness of unconventional policy tools, aimed at achieving both price and financial stability, as they reveal two important findings. First, since the implementation of QE, there has been an increase in bank lending activity. Second, this increase in bank loans is mostly due to the increase in the bank-to-bank lending level, rather than the bank-firm lending level. This fact is consistent with the process of progressive financialization of the economy that has been experienced in the last decades. This process, along with the evolution of markets as \textit{``markets for intermediaries rather than individuals or firms"} \cite{allen2001financial}, has implied that a large portion of financial interactions (via financial instruments) is intra-financial and only a small fraction of them is between financial system and real economy. This constitutes a particular concern in Europe, where the productive structure is mainly based on unlisted small and medium enterprises, which rely on the banking system, rather than on markets, to obtain funding. However, as in most serious debates, the truth lies somewhere in the middle. In fact, our explorative analysis also reveals that, since the implementation of APP, there has been an increase in both GDP and inflation. In a nutshell, our research has brought to light some important facts with regard to the effects of Quantitative Easing in the euro area. On the one hand, since the beginning of QE, there has not been a significant increase in bank-firm lending level, while, on the other hand, the overall euro area economy is experiencing growth and addressing the risk of deflation. 

\setcounter{secnumdepth}{0}
\section{Acknowledgement}
This work has been supported by the research project BigDataFinance. BigDataFinance project has received funding from the European Union's Horizon 2020 research and innovation programme under the Marie Sklodowska-Curie grant agreement No 675044.

\setcounter{secnumdepth}{0}
\bibliographystyle{ieeetr}
\bibliography{references}

\end{document}